\documentclass[letter,twocolumn]{jpsj3}
\usepackage{txfonts}

\title{Long-Period Charge Correlations in Charge-Frustrated Molecular $\theta$-(BEDT-TTF)$_2$X}
\author{Makoto Naka$^{\rm 1}$ and Hitoshi Seo$^{\rm 1,2,3}$}
\inst{$^{\rm 1}$RIKEN Center for Emergent Matter Science (CEMS), Wako, Saitama 351-0198 \\
$^{\rm 2}$Condensed Matter Theory Laboratory, RIKEN, Wako, Saitama 351-0198 \\
$^{\rm 3}$JST-CREST, Wako, Saitama 351-0198}
\abst{
Charge ordering (CO) with long-periodicities in a geometrically frustrated system, $\theta$-(BEDT-TTF)$_2$X, is numerically studied. 
We consider interacting fermion models on the anisotropic triangluar lattice: the extended Hubbard model and the spinless fermion model, taking account of the inter-site Coulomb interactions on not only the nearest neighbor (NN) but also the next nearest neighbor (NNN) bonds. 
By applying Lanczos exact diagonalization method, we find that long-period charge correlations become dominant when NNN terms are included, in the charge-frustrated region. 
The calculated optical excitation spectra show that the charge transfer excitations in the long-period CO 
extend to a lower energy range compared to those in the short-period CO states including the 3-fold CO stabilized when NNN terms are absent.
Our results suggest that the charge correlations experimentally observed in the title compounds near the CO insulating phase can be viewed as a tendency toward `tilted stripe'-type CO with large charge fluctuations. 
}
\kword{Charge Order, Geometrical Frustration, $\theta$-(BEDT-TTF)$_2$X, Optical Spectra, Lanzcos Exact Diagonalization, Extended Hubbard Model, Spinless Fermion Model}

\begin{document}
\maketitle
Molecular charge transfer salts have been studied extensively since they exhibit a variety of electronic phases due to the interplay of electron correlation effects and different kinds of  molecular packing.~\cite{fukuyama} 
In these compounds, charge order (CO) due to strong correlation is frequently observed; 
a periodic arrangement of localized electrons leads to the insulating state, whose spacial pattern determines the electronic properties such as transport phenomena and dielectric responses. 
In such systems, exotic features have been reported to emerge, e.g., giant nonlinear conduction~\cite{terasaki, inagaki}, ferroelectricity due to CO~\cite{monceau, yamamoto}, and glassy nature of charge correlation.~\cite{kagawa, chiba, nad} 

Quasi-two-dimensional compounds $\theta$-(BEDT-TTF)$_2$X are one of such CO system~\cite{hmori}, where BEDT-TTF is bis(ethylenedithio)tetrathiafulvalene and abbreviated as ET, and X takes different monovalent anions. 
Their crystal structure consists of alternating ET and X layers; ET molecules are arranged in the anisotropic triangular lattice, constructing the conducting layer, while X are closed shell. 
The molecular orbitals of ET form a quarter-filled band in terms of holes as a consequence of the charge transfer from X. 
By substituting X, different behaviors of CO and their fluctuations are observed. 
Typical examples are X$=$RbM(SCN)$_4$ (M$=$Zn, Co; abbreviated as Rb salts), where charge fluctuation characterized by the wave vector ${\bf q} = (2/3, k, 1/4)$ [$\equiv {\bf q}_1^{\rm Rb}$] found by X-ray scattering develops from high temperatures, 
and a first-order metal-to-insulator transition to the horizontal stripe-type CO state (denoted h-stripe) occurs at $200$ K.~\cite{mwatanabe} 
On the other hand, in X=CsM(SCN)$_4$ (M$=$Zn, Co; abbreviated as Cs salts), charge fluctuations with ${\bf q} = (2/3, k, 1/3)$ [$\equiv {\bf q}_{\rm 1}^{\rm Cs}$] and ${\bf q} = (0, k, 1/2)$, the latter owing to the h-stripe pattern within the two-dimensional plane, coexist below $120$ K.~\cite{nogami, mwatanabe2} 
The wave vectors ${\bf q}_{\rm 1}^{\rm Rb}$ and ${\bf q}_{\rm 1}^{\rm Cs}$ (we refer to them as ${\bf q}_{\rm 1}$ in the following) correspond to large unit cell sizes in real space, i.e., long-period charge correlations. 

To theoretically investigate electronic properties of $\theta$-(ET)$_2$X, effective models for interacting electrons on the anisotropic triangular lattice have been extensively studied.~\cite{seoreview}. 
Many examined the ground state properties of the extended Hubbard model (EHM), taking into account of the Coulomb interactions between the nearest neighbor (NN) ET molecules  along the bonds of the triangular lattice. 
When the high-temperature crystal structure is adopted, there are two independent bonds, and it has been shown that stripe-type CO insulating states are destabilized when NN Coulomb energies are comparable along the geometrical frustrated triangular bonds: charge frustration~\cite{merino, seoex}. 
There, a nonstripe-type CO metallic state called 3-fold CO~\cite{tmori, kaneko} is stabilized between the two stripe-type CO phases.~\cite{hwatanabe, nishimoto} 

However, the wave vectors for the CO and correlations observed in experiments do not agree with such calculations; additional ingredients are needed. 
As for the h-stripe, calculations taking into account of the lattice distortions seen experimentally reproduce its stability~\cite{seo, udagawa, tanaka, clay}, while other origins have been proposed~\cite{hotta, kuroki}. 
On the other hand, ${\bf q}_{\rm 1}$ charge fluctuation is frequently discussed~\cite{kaneko, hwatanabe, nishimoto, udagawa, tanaka, hotta, canocortes} in the context of the 3-fold CO, since it appears near the border of the stripe-type CO state; a situation analogous to experiments. 
The 3-fold CO pattern is also found in other studies for triangular lattice systems~\cite{motrunich, nagano}, indicating its strong commensurability to the triangular lattice. 
To reconcile the discrepancy from ${\bf q}_{\rm 1}$, the inclusion of longer distance Coulomb interactions has been pointed out.~\cite{tmori, kuroki}
Nevertheless, the nature of the long-period charge correlations is not elucidated: whether they are analogous to the 3-fold CO or not, and how they can be characterized. 

To address these issues, here we numerically investigate the stability of CO with long periodicities compared to short-period CO, i.e., stripe-type and 3-fold, and their optical response in order to characterize them. 
By using Lanczos exact diagonalization technique to the EHM and the spinless fermion model (SFM) on the anisotropic triangular lattice, we discuss the ground-state phase diagram in the presence of both the NN and further next nearest neighbor (NNN) Coulomb interactions. 
We will show that, once the NNN terms are turned on, long-period CO states are widely stabilized in the region between the stripe-type CO phases, instead of the 3-fold CO found in previous studies including only NN terms.
The optical absorption spectra indicate that the long-period CO show large charge fluctuations; they have a relatively small charge gap compared to the short-period CO. 
Finally, we discuss relevances of our results to the experimental results. 
%
%
%
\begin{figure*}[t]
\begin{center}
\includegraphics[width=2.0\columnwidth, clip]{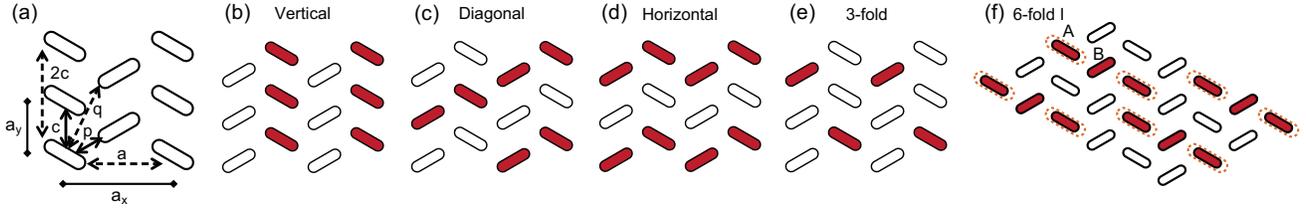}
\end{center}
\caption{(Color online) 
(a) Schematics lattice structure for $\theta$-(ET)$_2$X. 
Ellipses represent the ET molecules. 
Solid and broken arrows denote the NN and NNN bonds. 
Schematic charge configurations for the CO states are shown in (b) -- (f). 
The colored and white ellipses represent the charge-rich and charge-poor sites. 
In (f) 6-fold I CO, the charge-rich sites surrounded by the broken ellipses belong to the sublattice A, and the other charge-rich sites belong to B. 
}
\label{fig:lattice}
\end{figure*}
%
%
%

The EHM is described by the Hamiltonian defined as 
\begin{align}
{\cal H}_{\rm EHM} 
&=  \sum_{\langle ij \rangle \sigma} t_{ij}  (c_{i \sigma}^{\dagger} c_{j \sigma} + {\rm H.c.}) \notag \\
&+ U \sum_{i} n_{i \uparrow} n_{i \downarrow} + \sum_{\langle ij \rangle} V_{ij} n_{i} n_{j}, 
\end{align}
where $c_{i \sigma}$ is an annihilation operator for a hole at the $i$-th site with spin $\sigma (=\uparrow, \downarrow)$, 
and $n_{i} (\equiv \sum_{\sigma} n_{i \sigma} \equiv \sum_{\sigma} c_{i \sigma}^{\dagger} c_{i \sigma})$ is a number operator. 
We introduce the transfer integrals ($t_{ij}$), the on-site Coulomb interaction ($U$), and inter-site Coulomb interactions ($V_{ij}$). 
The hole picture is taken so that the system is quarter-filled. 

The SFM neglecting the spin degree of freedom in EHM is an effective model approximately describing the large $U$ limit. 
The Hamiltonian is written as 
\begin{align}
{\cal H}_{\rm SFM} 
= \sum_{\langle ij \rangle} t_{ij}  (f_{i}^{\dagger} f_{j } + {\rm H.c.})
+ \sum_{\langle ij \rangle} V_{ij} \tilde{n}_{i} \tilde{n}_{j}, 
\end{align}
where $f_{i}$ is an annihilation operator for a spinless fermion at the $i$-th site, 
and $\tilde{n}_{i} \equiv f^{\dagger}_{i} f_{i}$ is a number operator. 
Quarter-filling in the EHM corresponds to half-filling in the SFM, i.e., one fermion per two sites. 

We calculate the charge correlation function, $N({\bf q}) = \frac{1}{N^2} \sum_{ij} \langle 0 | n_{i} n_{j} | 0 \rangle e^{ - i  {\bf q} \cdot ({\bf R}_{i} - {\bf R}_{j}) }$ for the EHM, and $N({\bf q}) = \frac{1}{N^2} \sum_{ij} \langle 0 | \tilde{n}_{i} \tilde{n}_{j} | 0 \rangle e^{ - i  {\bf q} \cdot ({\bf R}_{i} - {\bf R}_{j}) }$ for the SFM, where ${\bf R}_{i}$ is the position vector of the $i$-th site and N is the number of sites in the finite-size cluster. 
The charge dynamics are examined by calculating the optical absorption spectra defined by 
\begin{align}
\alpha _\xi (\omega) = 
- \frac{e^2}{N} 
\rm{Im} \langle 0 | j_{\xi} \frac{1}{\omega - {\cal H} + E_{0} + i \eta} j_{\xi} | 0 \rangle, 
\end{align}
where  $\xi$ take two-dimensional Cartesian coordinates, $| 0 \rangle$ and $E_{0}$ are the ground-state wave function and energy, respectively, and $\eta$ is a broadening factor. 
The current operator is defined as ${\bf j} = i \sum_{\langle ij \rangle \sigma}  t_{ij} ({\bf R}_{i} - {\bf R}_{j}) (c^{\dagger}_{i \sigma} c_{j \sigma} - c^{\dagger}_{j \sigma} c_{i \sigma})$ for the EHM, and ${\bf j} = i \sum_{\langle ij \rangle}  t_{ij} ({\bf R}_{i} - {\bf R}_{j}) (f^{\dagger}_{i} f_{j} - f^{\dagger}_{j} f_{i})$ for the SFM. 

The two-dimensional arrangement of molecular sites and notations of the bonds, where interactions are taken into account, are shown in Fig.~\ref{fig:lattice} (a). 
$a_{x}$ and $a_{y}$ represent the lattice constants; we fix them as $\sqrt{3}$ and $1$, respectively. 
$\rm p$ and $\rm c$ are the NN bonds, while we refer to $\rm a$, $\rm q$, and $\rm 2c$ as the NNN bonds. 
According to previous works without the NNN terms,~\cite{tmori, seo, merino, kaneko, hwatanabe, nishimoto, hotta} it is known that the vertical stripe-type (v-stripe), diagonal stripe-type (d-stripe) and 3-fold CO, schematically shown in Fig.~\ref{fig:lattice} (b), (c), and (e), are stabilized for $V_{\rm p} \gg V_{\rm c}$, $V_{\rm c} \gg V_{\rm p}$ and $V_{\rm p} \sim V_{\rm c}$ in the EHM when $U$ is large, respectively, while d-stripe phase is replaced by h-stripe [Fig.~\ref{fig:lattice} (d)] in the SFM. 
In this study, in addition to these short-period CO, we explore the possibility of charge correlations with longer periodicities seen in experiments. 
We perform a comprehensive analysis of the CO states by the complimentary use of EHM and SFM, which enables us to choose different cluster sizes available for the Lanczos exact diagonalization method, especially in the latter. 
In the following, we show results for the EHM with a 18-site cluster, and investigate the competition between the 3-fold CO and long-period CO; a drawback is that in this case all the stripe-type CO cannot be included. 
To investigate such a competition between the different short-period CO and long-period CO, we consider the SFM with a 24-site cluster; the cluster that we use cannot include the 3-fold CO state but a similar pattern can be taken. 
We note that in the 3-fold CO in EHM the occupancy of the charge rich sites becomes larger than one in the large-$V_{ij}$ limit, therefore cannot be described in the SFM. 

We set $t_{\rm c} = 0$ for simplicity; this is often chosen to describe the case for the Cs salts, where an estimation by the extended H\"{u}ckel method gives $| t_{\rm c} / t_{\rm p} | < 0.1$.~\cite{mwatanabe2} 
The ratios of the inter-site Coulomb interactions are also estimated:~\cite{tmori} 
the NN energies, $V_{\rm p}$ and $V_{\rm c}$, are nearly equal, and the three kinds of NNN energies, $V_{\rm a}$, $V_{\rm q}$, and $V_{\rm 2c}$, are almost comparable with each other, and about $50 \%$ of the NN interactions. 
We fix $U = 2$ (for EHM) and $V_{\rm p} = 1$, as a unit of energy, and vary $V_{\rm c}$ to control the degree of charge frustration. 
As for the NNN terms, we set their ratios as $V_{\rm a} \colon V_{\rm q} = V_{\rm NNN} \colon 0.8 V_{\rm NNN}$ for the EHM ($V_{\rm 2c}$ cannot be taken into account because of the cluster shape), and $V_{\rm a} \colon V_{\rm q} \colon V_{\rm 2c} = V_{\rm NNN} \colon V_{\rm NNN} \colon 1.2 V_{\rm NNN}$ for the SFM, where $V_{\rm NNN}$ is controlling parameter for the strength of the NNN Coulomb interaction. 
$t_{\rm p}$ is set to $0.2$ for the calculations of $N({\bf q})$, which is typically chosen in the literature ($U/t_{\rm p} = 10$), while we vary it for $\alpha_\xi(\omega)$ in order to identify the origin of the excitations. 

%
%
%
\begin{figure}[t]
\begin{center}
\includegraphics[width=0.7\columnwidth, clip]{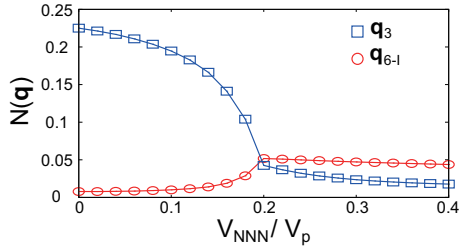}
\end{center}
\caption{(Color online) 
$V_{\rm NNN}$-dependence of the charge correlation functions $N({\bf q})$ for ${\bf q} = {\bf q}_{\rm 3}$ and ${\bf q}_{\rm 6-I}$ (see text) for the EHM. 
The other parameters are fixed as $U = 2$, $V_{\rm p} = 1$, $t_{\rm p} = 0.2$ and $V_{\rm c} = 1.4$. 
}
\label{fig:cc_EHM}
\end{figure}
%
%
%
\begin{figure}[t]
\begin{center}
\includegraphics[width=0.7\columnwidth, clip]{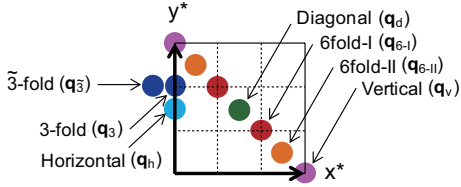}
\end{center}
\caption{(Color online) 
Wave vectors of the short-period and long-period CO in the $1$-st Brillouin zone. 
}
\label{fig:kspace}
\end{figure}
%
%
%
First, we discuss ground state properties of the EHM, when the system is in the 3-fold CO phase without the NNN terms ($V_{\rm c} = 1.4V_{\rm p}$).
In Fig.~\ref{fig:cc_EHM}, the $V_{\rm NNN}$-dependence of the charge correlation function $N({\bf q})$ for ${\bf q}_{\rm 3} \equiv (0, 2/3)$ and ${\bf q}_{\rm 6-I} \equiv (2/3, 1/3)$ is shown, taking the largest among other ${\bf q}$. 
${\bf q}_{\rm 3}$ and ${\bf q}_{\rm 6-I}$ represent the wave vectors for the 3-fold CO and a long-period wave vector which we call here 6-fold I CO, whose real space charge configuration, containing 6 sites in the unit cell, is schematically illustrated in Fig.~\ref{fig:lattice} (f); note that this pattern has not been considered in ref.~\citen{tmori}. 
When we increase $V_{\rm NNN}$, $N({\bf q}_{\rm 3})$ decreases rapidly and $N({\bf q}_{\rm 6-I})$ becomes the largest for $V_{\rm NNN} \gtrsim 0.2 V_{\rm p}$. 
This indicates that the 3-fold CO is destroyed and the 6-fold I CO is stabilized instead, in the presence of rather small NNN Coulomb interactions. 

%
%
%
\begin{figure}[t]
\begin{center}
\includegraphics[width=0.9\columnwidth, clip]{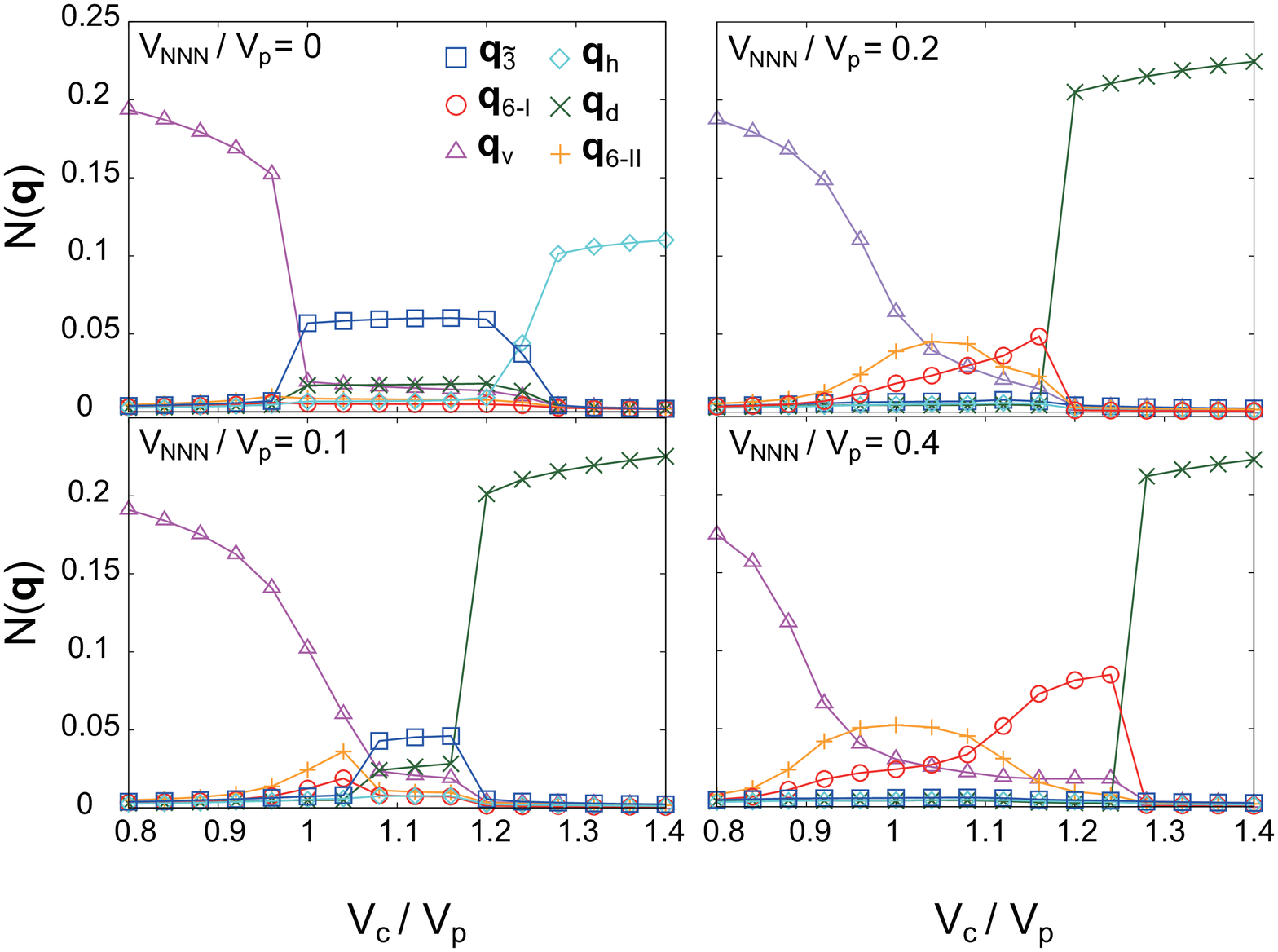}
\end{center}
\caption{(Color online) 
$V_{\rm c}$-dependence of the charge correlation functions $N({\bf q})$ for the SFM for the wave vectors discussed in the text. 
Parameters are set as $t_{\rm p} = 0.2$, $V_{\rm p} = 1$, and $V_{\rm NNN} = 0$ (left upper panel), $0.1$ (left lower panel), $0.2$ (right upper panel), and $0.4$ (right lower panel). 
}
\label{fig:cc_SFM}
\end{figure}
%
%
%
%
%
%
\begin{figure}[t]
\begin{center}
\includegraphics[width=0.7\columnwidth, clip]{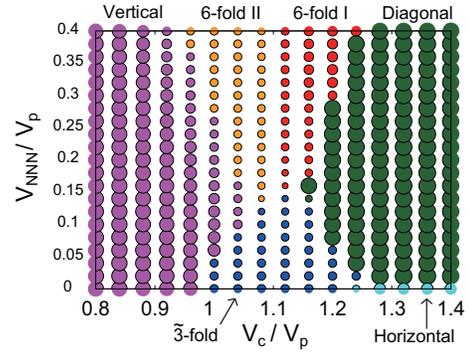}
\end{center}
\caption{(Color online) 
Ground-state phase diagram of the SFM. 
The area of the circles represents the amplitude of $N({\bf q}_{\rm max})$, where ${\bf q}_{\rm max}$ is the wave vector where $N({\bf q})$ shows the largest value. 
}
\label{fig:pd_SFM}
\end{figure}
%
%
%
The SFM also shows the stability of long-period CO when NNN interactions are included. 
The $V_{\rm c}$-dependence of the charge correlation function is shown in Fig.~\ref{fig:cc_SFM}, for different values of $V_{\rm NNN}$. 
Now we can compare different stripe-type CO as mentioned above. 
The v-, d-, and h-stripe are characterized by ${\bf q}_{\rm v} \equiv (1, 0)$, ${\bf q}_{\rm d} \equiv (1/2, 1/2)$ and ${\bf q}_{\rm h} \equiv (0, 1/2)$, and a long-period CO other than 6-fold I, which we call 6-fold II CO, is characterized by ${\bf q}_{\rm 6-II} \equiv (5/6, 1/6)$. 
Although the 3-fold CO cannot be included in this analysis due to the boundary condition, a nonstripe `$\tilde{\rm 3}$-fold' CO characterized by ${\bf q}_{\tilde{\rm 3}} \equiv (-1/6, 2/3)$, whose wave vector is close to ${\bf q}_{\rm 3}$, is taken into account. 
In the absence of $V_{\rm NNN}$, $N({\bf q}_{\rm v})$ and $N({\bf q}_{\rm h})$ have large values in the small and large $V_{\rm c}$ region, respectively, while ${\bf q}_{\tilde{\rm 3}}$ becomes the largest in the medium $V_{\rm c}$ region; the behavior is the same with results on a 24-site cluster but with a different shape if we replace $\tilde{\rm 3}$-fold with 3-fold CO~\cite{hotta}, therefore ${\bf q}_{\tilde{\rm 3}}$ can be considered analogous to the 3-fold state. 
When we introduce a small amount of $V_{\rm NNN}$, 
the dominant CO immediately changes from h- to d-stripe in the large $V_{\rm c}$ region, 
and the $\tilde{\rm 3}$-fold CO region becomes narrower. 
By further increasing $V_{\rm NNN}$, in the medium $V_{\rm c}$ region, the long-period charge correlation functions $N({\bf q}_{\rm 6-I})$ and $N({\bf q}_{\rm 6-II})$ develop, and $N({\bf q}_{\tilde{\rm 3}})$ is strongly suppressed. 
In Fig.~\ref{fig:pd_SFM}, we show the ground-state phase diagram from these results. 
One can see that the 6-fold I and II CO are stabilized by $V_{\rm NNN}$ between the v- and d-stripe phases, by relatively small amount of NNN interactions. 

Through these analyses, we find that the long-period CO appear in the presence of the NNN Coulomb interactions strongly suppressing the short-period CO; this is because they act on the bonds between the charge rich sites in the stripe-type and 3-fold CO.~\cite{tmori} 
It is noteworthy that the wave vectors ${\bf q}_{\rm 6-I}$ and ${\bf q}_{\rm 6-II}$ are located between ${\bf q}_{\rm v}$ and ${\bf q}_{\rm d}$ in the reciprocal space, as shown in Fig.~\ref{fig:kspace}. 
On the other hand, ${\bf q}_{\rm 3}$ and ${\bf q}_{\tilde{\rm 3}}$ locate in a separated area. 
These imply that the long-period CO can be considered as a stripe-type charge modulation, rather than an analog of the 3-fold states, as can also be seen in the real space patterns in Fig.\ref{fig:lattice} (b) -- (f). 
Such characters also appear in the charge dynamics discussed in the following. 

%
%
%
\begin{figure}[t]
\begin{center}
\includegraphics[width=0.7\columnwidth, clip]{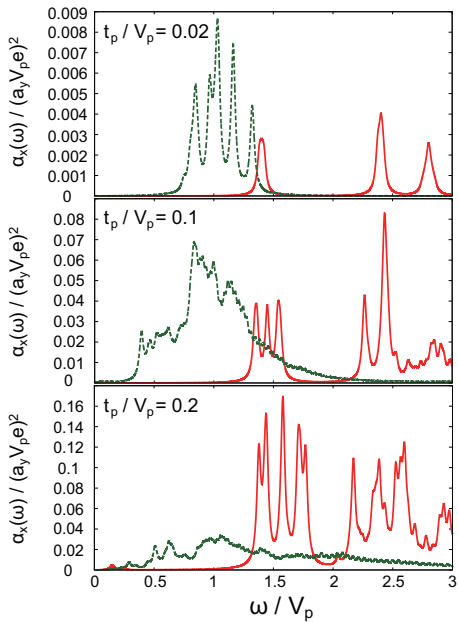}
\end{center}
\caption{(Color online) 
Optical absorption spectra in the 3-fold CO (solid line) and the 6-fold I CO (broken line) for $t_{\rm p} = 0.02$ (upper panel), $0.1$ (middle panel), and $0.2$ (lower panel) for the EHM. 
Other parameters are chosen as $U = 2$, $V_{\rm p} = 1$, and $V_{\rm c} = 1.4$. 
}
\label{fig:opt_EHM}
\end{figure}
%
%
%
We compare the optical absorption spectra for the 3-fold and 6-fold I CO in the EHM, presented in Fig.~\ref{fig:opt_EHM}. 
In the 3-fold CO, three peaks are seen in the absorption spectrum for $t_{\rm p} = 0.02$. 
These peaks can be assigned to the charge transfer (CT) excitations from the charge-rich sites of the 3-fold CO. 
Their CT excitation energies are given as $2V_{\rm p} + V_{\rm c} - U$,  $3V_{\rm p} + V_{\rm c} - U$, $V_{\rm p} + V_{\rm c}$, and $2V_{\rm p} + 2V_{\rm c} - U$ in the classical limit; the second and third are degenerate in this parameter set. 
When we increase $t_{\rm p}$ toward realistic values, one can see that these CT excitation peaks split, but the charge excitation gap is almost unchanged~\cite{nishimoto}. 
On the other hand, in the 6-fold I CO, the absorption spectrum is roughly divided into low energy and high energy peaks at $t_{\rm p} = 0.02$. 
The former peaks are attributed to the CT excitations from the sublattice A in the 6-fold I CO [see Fig.\ref{fig:lattice} (f)], whose excitation energies are given as $- V_{\rm c} + 0.6V_{\rm NNN} + U$ and $V_{\rm p}$ in the classical limit. 
The high energy peaks can be assigned by the excitations from the sublattice B,  given as $- 2V_{\rm p} + V_{\rm c} - 0.6V_{\rm NNN} + U$ and $- V_{\rm p} + 2V_{\rm c}  - 1.2V_{\rm NNN}$. 
When we increase the value of $t_{\rm p}$, these CT excitation peaks are remarkably broadened 
toward a continuous spectrum, and the charge excitation gap becomes appreciably small. 
From these results, we can conclude that the 6-fold I CO has a strongly fluctuating nature, in sharp contrast with the 3-fold CO state which is robust due to its matching with the triangular lattice. 
Such a feature is also seen in the charge correlation functions: 
$N({\bf q}_{\rm 6-I})$ is reduced by $21 \%$ due to the increase of the electron transfer $t_{\rm p}$ from $0.02$ to $0.2$, whereas the reduction in $N({\bf q}_{\rm 3})$ is only about $10 \%$. 

%
%
%
\begin{figure}[t]
\begin{center}
\includegraphics[width=0.7\columnwidth, clip]{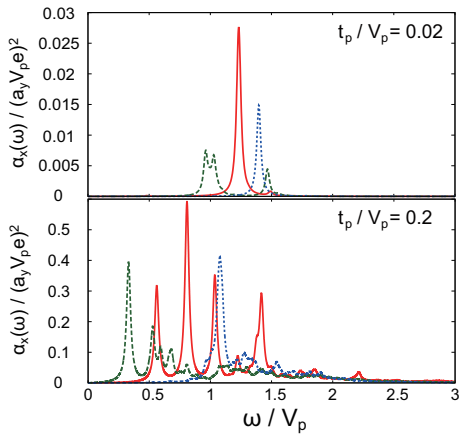}
\end{center}
\caption{(Color online) 
Optical absorption spectra in the vertical stripe-type ($V_{\rm c} = 0.8$, solid line), 6-fold I ($V_{\rm c} = 1.16$, broken line), and diagonal stripe-type ($V_{\rm c} = 1.28$, dot line) CO states for $t_{\rm p} = 0.02$ (upper panel) and $t_{\rm p} = 0.2$ (lower panel) for the SFM. 
We fix $V_{\rm p} = 1$ and $V_{\rm NNN} = 0.4$. 
}
\label{fig:opt_SFM}
\end{figure}
%
%
%
The optical absorption spectra of the SFM for the 6-fold I CO, v- and d-stripe are compared in Fig.~\ref{fig:opt_SFM}. 
In the stripe-type CO, single peak structures are seen for $t_{\rm p} = 0.02$; these peaks can be assigned to the CT excitations from the charge-rich sites, whose excitation energies in the classical limit are $3V_{\rm p} - 2V_{\rm c} - 0.4V_{\rm NNN}$ and $- V_{\rm p} + 2V_{\rm c} - 0.4V_{\rm NNN}$ for the v- and d-stripe, respectively. 
On the other hand, in the 6-fold I CO, the low and high energy peak structures similar to the case of the EHM are seen for $t_{\rm p} = 0.02$. 
Adopting the estimations above for the EHM and setting $U \to \infty$, the peaks can be attributed to the CT excitations from the sublattices A and B given by $V_{\rm p}$ and $- V_{\rm p} + 2V_{\rm c} + 0.4V_{\rm NNN}$, respectively. 
A common feature for all the three CO states here is that by increasing $t_{\rm p}$, the CT excitation peaks split into a wide energy range and the charge excitation gaps become reduced. 
This is another reason why we consider the long-period states as a stripe-type CO. 
A characteristic feature for the 6-fold I CO compared to the v- and d-stripe is the smallest charge gap, both in the classical limit and when $t_{\rm p}$ is included. 
These results suggest that the low-energy CT excitations seen in our calculations show more fluctuations in the long-period CO state due to the `tilting' of the stripes [see Fig.\ref{fig:lattice} (f)] compared to the unfrustrated v- and d-stripe states aligned along the crystal axes. 

Finally, let us compare our results with experiments for $\theta$-(ET)$_2$X. 
In the Rb and Cs salts, as mentioned above, the ratio of the bare NN and NNN Coulomb interactions are roughly estimated by using results of the extended H\"{u}ckel calculations as $V_{\rm c} \sim 1$ -- $1.3 V_{\rm p}$ and $V_{\rm NNN} \sim 0.5 V_{\rm p}$.~\cite{tmori} 
In the phase diagram of the SFM in Fig.~\ref{fig:pd_SFM}, 
they correspond to the long-period 6-fold I, 6-fold II, or d-stripe CO phase. 
Among them, the wave vector ${\bf q}_{\rm 6-I}$ is identical to ${\bf q}_{\rm 1}^{\rm Cs}$, i.e., the long-period charge correlation observed in the Cs salts. 
We performed the analysis of the SFM on another 24-site cluster where the 6-fold I and II CO do not fit the boundary condition. 
In this case, we obtain a long-period CO characterized by the wave vector ${\bf q} = (3/4, 1/4)$, which is close to ${\bf q}_{\rm 1}^{\rm Rb}$. 
As for the charge dynamics, our results show that the long-period CO have low energy charge excitations compared to the short-period CO. 
This is qualitatively consistent with the experimental results that the Cs salts, where the long-period and h-stripe charge correlations coexist, show lower energy optical weight than the Rb salts where only h-stripe is present at low temperatures~\cite{tajima, wang, suzuki, hashimoto}. 
The low energy spectral weight can be attributed to the CT excitation with large fluctuations originated from the tilted stripe-type charge correlation. 

As above mentioned, the 3-fold CO is often referred to the origin of the ${\bf q}_{\rm 1}$ CO in the experiments. 
However, in this study, it is found that the 3-fold CO is strongly suppressed and replaced by long-period CO in the presence of relatively small values of NNN Coulomb energies, and the 3-fold CO has a relatively rigid charge gap; it would give rise to superlattice reflections in the X-ray diffraction image and a clear charge excitation gap. 
However, the observations actually show a diffusive nature of the diffraction pattern and the low energy charge excitation in the ${\bf q}_{\rm 1}$ CO. 
Another point is that the anisotropy in the optical spectra, along the crystallographic axes in the two-dimensional plane, are distinct between the 3-fold and long-period CO states; 
the former shows large anisotropy~\cite{nishimoto} while it is nearly isotropic in the latter due to the tilting of the stripe pattern, which is in fact consistent with recent measurements~\cite{hashimoto}. 

We note that the wave vectors of the long-period CO between v- and d-stripes depend on the size and the shape of the cluster which we use in the numerical calculations. 
Nevertheless, they are on the line connecting ${\bf q}_{\rm v}$ and ${\bf q}_{\rm d}$; therefore, our discussion here that long-period CO are essentially tilted stripe CO states holds in general. 
If we can perform an analysis in the thermodynamic limit, we can deduce that the wave vector of the long-period CO continuously varies from ${\bf q}_{\rm v}$ to ${\bf q}_{\rm d}$ crossing the phase diagram. 
Whether a devil's staircase-type successive phase transition occurs or an incommensurate CO appear remains as an open question. 
The origin of the glassy features found in the Cs salts~\cite{nad, chiba} as well as in rapidly-cooled Rb salts~\cite{kagawa} is still elusive, while an analysis based on EHM without frustration suggests a possible origin due to phase separation near the CO phase~\cite{yoshimi}. 
Such a scenario was actually first introduced in the context of the stripe phase in transition metal oxides~\cite{schmalian}. 
The suggestion here is that the fluctuating stripes are actually already observed in the molecular compounds, 
then its relation with the glassy nature will be an interesting future issue. 

The authors would like to thank K.~Hashimoto, S.~Ishihara, F.~Kagawa, H.~Nakano, and T.~Sasaki for valuable discussions. 
This material is based upon work supported in part by Grant-in-Aid for Scientific Research (No. 24108511) from MEXT.

\end{document}